\documentstyle[12pt]{article}
\begin{document}

\baselineskip=24pt plus 2pt
\hfill\hbox{NCKU-HEP/96-03}
\begin{center}

{\large \bf On the energy of a charged dilaton black hole}\\
\vspace{5mm}
I-Ching Yang $^\dag$ \footnote{E-mail:icyang@ibm65.phys.ncku.edu.tw},
Ching-Tzung Yeh $^{\dag\ddag}$ ,\\
Rue-Ron Hsu $^\dag$ \footnote{E-mail:rrhsu@mail.ncku.edu.tw}
and~ Chin-Rong Lee $^\S$ \footnote{E-mail:phycrl@ccunix.ccu.edu.tw}
\vspace{5mm}

$^\dag$Department of Physics, National Cheng Kung University \\
Tainan, Taiwan 701, Republic of China \\
$^\ddag$Department of Physics, National Central University \\ 
Chung-Li, Taiwan 32054, Republic of China \\
$^\S$Department of Physics, National Chung Cheng University \\
Chia-Yi, Taiwan 62117, Republic of China \\

\end{center}
\vspace{5mm}

\begin{center}
{\bf ABSTRACT}
\end{center}

    Employing energy-momentum pseudotensor of Einstein, we obtian the
energy distribution  of a dyonic dilaton black hole. The energy distribution
of this black hole depends on mass $ M $, electric charge
$ Q_{e} $, magnetic charge $ Q_{m} $ and asymptotic value of the dilaton
$ \phi_{0} $. We also make some comparisons between the  
results of Virbhadra et. al. and ours.

\vspace{2mm} 
\noindent
{PASC:04.20.-q, 04.50,+h}
\newpage

    Charged dilaton black hole solutions have been
obtained by Grafinkle, Horowitz and Strominger(GHS solutions)~\cite{1}.
They are static spherical symmetric black hole solutions of the dilaton
gravity theory. In this theory the gravity is coupled to electromagnetic
and dilaton fields and can be describled by the four-dimensional 
effective string action. The action can be expresses as 
\begin{equation}
  I = \int d^4 x \sqrt{-g}  \left[ -R + 2(\nabla \phi)^2 + e^{-2\phi}F^2 \right] .
\end{equation} 
In a peculiar coordinate form, GHS solutions have  a singularity whose area is 
zero at  \( r^\ast = \frac {Q_{m}^2}{M} e^{2\phi_{0}} \) .
To characterize a gravitational point source, their choice of the coordinate system may not be the most appropriate one. Thus, Cheng, Lin and Hsu~\cite{2} use the standard spherical coordinate system which is more suitable for describling the structure of 
the charged dilaton black hole. They obtained the more general solutions which named the dyonic dilaton black hole solutions.
GHS solutions were found to be the special cases of dyonic dilaton black hole solutions when electric or
magnetic charges are switched off. For these special cases, two solutions are related by some 
coordinate transformations~\cite{2}. For example, 
The singularity \( r^\ast = \frac {Q_{m}^2}{M} e^{2\phi_{0}} \) of GHS solutions 
corresponds to the singularity \( r = 0 \) of the dyonic dilaton black hole solutions.   

    Recently, Virbhadra et. al.~\cite{3} get the gravitational energy of charged 
dilaton black hole solutions. Based on the GHS solutions, they found a charge independent result 
\begin{equation}
   E(r) = M ,
\end{equation}
in which the positive energy is confined to the interior of the black hole. Their result is different from
the gravitational energy of Reissner-Nordstr${\ddot{o}}$m (RN) solutions~\cite{4} 
\begin{equation}
   E(r) = M - \frac{Q^2}{2r} ,
\end{equation}
where the energy is shared by the interior as well as the exterior of the black hole and becomes
negative for \( r < \frac{Q^2}{2M} \).
Moreover, Chamorro and Virbhadra~\cite{5}~\cite{6} considered a generalized action
\begin{equation}
  I = \int d^4 x \sqrt{-g}  \left[ -R + 2(\nabla \phi)^2 + e^{-2\alpha\phi}F^2 \right] ,
\end{equation} 
where \( \alpha \) is a dimensionless parameter which controls the coupling
between the dilaton and the Maxwell fields. They found the gravitational
energy of charged black hole is 
\begin{equation}
  E(r) = M -\frac{Q^2}{2r} \left( 1 - \alpha^2 \right) .
\end{equation}
The energy remains positive for all \( r \) as long as \( \alpha^2 > 1 \), and the
total energy is also shared by both regions.

    In this paper, we will investigate the gravitational energy of dyonic dilaton black hole
in the standard spherical coordinate instead. We will also make  some comments on the results of Virbhadra et. al. and ours.

    The dyonic dilaton black holes are static, spherical symmetric 
solutions of the dilaton gravity theory.
In terms of the standard spherical coordinate~\cite{2}, the dyonic dilaton black hole solutions are
\begin{eqnarray}
  ds^2  & = & \Delta^2 dt^2 - \frac{\sigma^2}{\Delta^2} dr^2 - r^2 d\theta^2 - r^2 \sin^2\theta d\varphi^2 , \\
  \sigma ^2 & = & \frac{r^2}{r^2 + \lambda^2} , \\
  \Delta ^2 & = & 1 - \frac{2M}{r^2} \sqrt{r^2 + \lambda^2} + \frac{\beta}{r^2} , \\
  \lambda & = & \frac{1}{2M} \left(Q_{e}^2 e^{2\phi_{0}} - Q_{m}^2 e^{-2\phi_{0}} \right) , \\
  \beta & = & Q_{e}^2e^{2\phi_{0}} + Q_{m}^2e^{-2\phi_{0}} ,\\
e^{2\phi} & = & e^{-2\phi{0}} \left( 1 - {2\lambda\over {\sqrt{r^2 + \lambda^2}+\lambda}} \right) ,\\
  F_{01} & = & {Q_e \over r^2}e^{2\phi} ,\\
  F_{23} & = & {Q_m \over r^2} .
\end{eqnarray}
The properties of the dyonic dilaton black holes are characterized by
mass \( M \), electric charge \(Q_{e} \), magnetic charge \(Q_{m} \)
and asymptotic value of the dilaton \( \phi_{0} \). The structures of
the dyonic dilaton black holes are similar to that of the RN ~\cite{2} 
black holes.

    The well known energy - momentum complex of Einstein~\cite{6} 
is defined as
\begin{equation}
  \Theta_{\mu}^{\nu} = \frac{1}{16\pi} \frac{\partial H_{\mu}^{\nu\sigma}}{\partial x^{\sigma}} , 
\end{equation}                                   
where
\begin{equation}
  H_{\mu}^{\nu\sigma} = \frac {g_{\mu\rho}}{\sqrt{-g}} \frac{\partial}{\partial x^{\eta}} \left[\left( -g \right) \left( g^{\nu\rho} g^{\sigma\eta} - g^{\sigma\rho} g^{\nu\eta} \right) \right] .
\end{equation}
The Greek indices run from 0 to 3 and \(x^{0} \) is the time coordinate.
Then, the energy cpmponent \( E \) is given by 
\begin{eqnarray}
  E & = & \int \int \int \Theta_{0}^{0} dx^{1} dx^{2} dx^{3} \nonumber \\
    & = & \frac {1}{16\pi} \int \int \int \frac{\partial H_{0}^{0l}}{\partial x^{l}} dx^{1} dx^{2} dx^{3} , 
\end{eqnarray}
where the Latin index takes values from 1 to 3.

    We carry out the energy component calculation in the quasi-Cartesian coordinate \(\left( t, x, y, z \right) \).           
The conversion of spherical coordinate into Cartesian coordiante is
\begin{equation}
   \left\{\begin{array}{l}
   x  =  r\sin \theta \cos \varphi \\
   y  =  r\sin \theta \sin \varphi \\
   z  =  r\cos \theta  
   \end{array}  ,
 \right .
\end{equation}
and it transforms the line element (6) into 
\begin{equation}
  ds^2 = \Delta^2 dt^2 - (dx^2 + dy^2 + dz^2 ) - \frac {\sigma^2 / \Delta^2 - 1 }{r^2}(xdx + ydy +zdz )^2 .
\end{equation}
Thus, we obtain the required components \(H_{0}^{0l} \) in Eq.(16),
\begin{eqnarray}
   H_{0}^{01} & = & \frac{2x\sigma}{r^2} \left( 1 - \frac{\Delta^2}{\sigma^2} \right) , \\
   H_{0}^{02} & = & \frac{2y\sigma}{r^2} \left( 1 - \frac{\Delta^2}{\sigma^2} \right) , \\
   H_{0}^{03} & = & \frac{2z\sigma}{r^2} \left( 1 - \frac{\Delta^2}{\sigma^2} \right) .
\end{eqnarray} 

    After plugging Eqs.(7),(8) and (19)-(21) into (16), and applying the Gauss
theorem, we evaluate the integral over the surface of a sphere with radius \( r \).
\begin{eqnarray}
   E(r) & = & \frac {1}{16\pi} \int \int \int \frac{\partial H_{0}^{0l}}{\partial x^{l}} dx dy dz \nonumber \\
        & = & \frac {1}{16\pi} \oint \frac{2\sigma}{r} \left( 1 - \frac{\Delta^2}{\sigma^2} \right) r^2 \sin \theta d \theta d \varphi .
\end{eqnarray}
Finally, we find the energy within a sphere with radius \( r \) is 
\begin{equation}
   E(r) = M + \frac{M\lambda^2}{r^2} - \frac{1}{2\sqrt{r^2 + \lambda^2}} \left[ \frac{\beta\lambda^2}{r^2} + \lambda^2 + \beta \right] .
\end{equation} 
The energy is shared both by the interior and by the exterior of the black hole. We plot the 
energy distributions of the dyonic black holes or the extremal dyonic black holes
by "GNUPLOT". For the dyonic black hole or the extremal dyonic black hole,
see Fig. 1 and Fig. 4, we find the energy distribution can be 
positive or negative. However, they are both positive in the region
\( r > r_{H} \). For the pure electric or pure magnetic charged black 
hole, ie \( Q_{e} = 0 \) or \( Q_{m} = 0 \), we find the remarkable
property that the energy distribution is always positive except at
singular point \( r = 0 \), see Fig. 2,3,5,6. 

Here, we note that
we can reparametrize \( Q_{e} \) and \( Q_{m} \) by new parameters
\( Q\) and \( \overline{Q} \),
\begin{eqnarray}
   Q & = & \frac{1}{\sqrt{2}} \left( Q_{e} + Q_{m} \right) \nonumber \\
   \overline{Q} & = & \frac{1}{\sqrt{2}} \left( Q_{e} - Q_{m} \right),
\end{eqnarray}
After this reparametrization, Eqs.(23), will give the same energy distribution of RN black hole,
as \( \phi_{0} = 0 \) and \( \overline{Q} = 0 \)~\cite{4}.
The correspondence of the energy distribution between the dyonic dilaton
black hole solutions and RN solutions can be understood by observing the
action Eqs.(1) or the solutions Eqs.(6)-(13). When dilaton field was 
suppressed, that is $ \phi_{0} = 0 $ and $ \lambda = 0 $ (or $ \overline{Q} = 0 $),
the dilaton gravity, Eqs.(1), will reduce to Einstein - Maxwell theory 
and the dyonic dilaton black hole solutions will become to be RN solutions.
Again, the results are same as the case of Schwarzschild black holes,
as we switch off \( Q_{e} \), \( Q_{m} \) and \( \phi_{0} \)~\cite{7}. 
 
    As expected, the energy distribution of the dyonic dilaton
black hole, depends on those parameters \( M \), \( Q_{e} \), \( Q_{m} \)
and \( \phi_{0} \). They are different from the results of Virbhadra et. al.
The differences do come from the different chioces of coordinate representations
in which energy distribution were evaluated. It can be understood that the 
local gravitation energy is a component of 4 momentum vector and it will
change as different coordinate is chosen . It seems that different coordinates
will respect to the different reference points of energy distribution.
However, if we are only interested in the ADM mass~\cite{8}
- total energy which is the limiting value of a certain flux integral
as a spherical surface expands to spatial infinity, we find
\begin{equation}
   M_{ADM} = {E(r) \mid }_{r \rightarrow \infty} = M.
\end{equation}
The result is same as the result of Virbhadra. It means that the ADM
mass is independent of coordinate representation of the black hole.

\newpage
\begin{center}
{\bf Acknowledgements.} 
\end{center} 
We thanks Prof. K.S. Virbhadra and Prof. J.M. Nester for useful  
comments and discussions.
This work is supported in part by the National Science Council of the
Republic of China under grants NSC-85-2112-M006-005 and NSC-85-0208-M194-002.

\newpage
\begin{figure}[hp]
   \input{fig1.tex}
   \caption{ The energy distribution of dyonic black hole with $ \phi_{0}=0 $, $ M=2 $, $ Q_{e}=1 $ and $ Q_{m}=1 $.}
   \vspace{5mm}
   \input{fig2.tex}
   \caption{ The energy distribution of pure electric black hole with $ \phi_{0}=0 $, $ M=2 $, $ Q_{e}=1 $ and $ Q_{m}=0 $.}
\end{figure} 
\newpage
\begin{figure}[hp]
   \input{fig3.tex}
   \caption{ The energy distribution of pure magnetic black hole with $ \phi_{0}=0 $, $ M=2 $, $ Q_{e}=0 $ and $ Q_{m}=1 $.}
   \vspace{5mm}
   \input{fig4.tex}
   \caption{ The energy distribution of extremal dyonic black hole with $ \phi_{0}=0 $, $ M=2 $, $ Q_{e}=\sqrt{2} $ and $ Q_{m}=\sqrt{2} $.}
\end{figure}
\newpage
\begin{figure}[hp]
   \input{fig5.tex}
   \caption{ The energy distribution of extremal electrically charged solution with $ \phi_{0}=0 $, $ M=2 $, $ Q_{e}=2\sqrt{2} $ and $ Q_{m}=0 $.}
   \vspace{5mm}
   \input{fig6.tex}
   \caption{ The energy distribution of extremal magnetically charged solution with $ \phi_{0}=0 $, $ M=2 $, $ Q_{e}=0 $ and $ Q_{m}=2\sqrt{2} $.}
\end{figure}

\end{document}